\documentclass[aps,twocolumn,showpacs,pre]{revtex4-1}
\usepackage{epsfig,amsmath,color,colordvi,verbatim}
\usepackage[normalem]{ulem}
\usepackage{braket}
\usepackage{hyperref}

\def\be{\begin{equation}}
\def\ee{\end{equation}}
\def\bee{\begin{eqnarray}}
\def\eee{\end{eqnarray}}

\def\with{\quad\mbox{with}\quad}

\begin{document}

\title{An efficient tool to calculate two-dimensional optical spectra 
for photoactive molecular complexes}

\author{Hong-Guang Duan$^{1,2}$, Arend G. Dijkstra$^{2}$, Peter
Nalbach$^{1,3}$} 
\author{Michael
Thorwart$^{1,3}$}\email{michael.thorwart@physik.uni-hamburg.de}
\affiliation{$^1$I.\ Institut f\"ur Theoretische Physik,  Universit\"at
Hamburg, Jungiusstra{\ss}e 9, 20355 Hamburg, Germany \\
$^2$Max Planck Institute for the Structure and Dynamics of Matter, Luruper
Chaussee 149, 22761, Hamburg, Germany\\
$^3$The Hamburg Centre for Ultrafast Imaging, Luruper Chaussee 149, 22761
Hamburg, Germany
}
\date{\today}

\begin{abstract}
We combine the coherent modified Redfield theory (CMRT) with the 
equation of motion-phase matching approach (PMA) to calculate two-dimensional 
photon echo spectra for photoactive molecular complexes with an intermediate 
strength of the coupling to their environment. Both techniques are highly 
efficient, yet they involve approximations at different levels. 
By explicitly comparing with the numerically exact quasi-adiabatic 
path integral approach, we show for the Fenna-Matthews-Olson 
complex that the CMRT describes the decay rates in the population dynamics well, 
but final stationary populations and the oscillation frequencies differ
slightly. In addition, we use the combined CMRT+PMA to calculate 
two-dimensional photon-echo spectra for a simple dimer model. 
We find excellent agreement with the exact path integral 
calculations at short waiting times where the dynamics is still 
coherent. For long waiting times, differences occur due to different final 
stationary states, specifically for strong system-bath coupling. 
For weak to intermediate system-bath couplings, which is most 
important for natural photosynthetic complexes,
the combined CMRT+PMA gives reasonable results 
with acceptable computational efforts.
\end{abstract}

%\pacs{87.15.H-, 03.65.Yz, 03.65.Ud, 03.65.Ta}
\maketitle

\maketitle

%%%%%%%%%%%%%%%%%%%%%%%%%%%%%%%%%%%%%%%%%%%%%%%%%%%%%%%%%% Intro
%%%%%%%%%%%%%%%%%%%%%%%%%%%%%%%%%%%%%%%%%%%%%%%%%%%%%%%%%%
\section{\label{sec:intro} Introduction}

Photosynthesis is the process used by plants and bacteria to convert 
the energy of sunlight into chemical energy in order to 
fuel the organism's activities. In the initial steps of the
photosynthetic process, pigment-protein complexes complete the
light-energy transfer and charge separation with a near unity quantum
efficiency. Light-harvesting molecular complexes achieve the energy transfer by
using an array of light-harvesting pigments that absorb the energy to 
form excitons and funnel the excitation to the reaction 
center \cite{Blankenship_Book}. To
investigate the energy transfer dynamics in the first steps of photosynthesis 
on the fs time scale, ultrafast spectroscopic tools are available. Among the
techniques used, two-dimensional (2D) photon echo spectroscopy is a powerful tool which
allows for direct mapping of the excitation energy pathways as a function of
absorption and emission wavelength \cite{2D_setups}. It is particularly
useful in examining photosynthetic systems in which the manifold of
electronic states is closely spaced and broadening through static disorder
yields highly congested spectra. Recent experimental 2D electronic
spectroscopic studies of the Fenna-Matthews-Olson (FMO) complex observed
coherent beating signals and, thus, raised interest in the interplay
between energy transfer, long-lived quantum coherence in photosynthetic
processes \cite{Engel-Nature-2007}, and low-frequency 
vibrations of the molecular back bone. Also in photoactive
marine cryptophyte algae \cite{Nature 463 644 (2010)}, the light-harvesting
complex LH2 \cite{Zigmantas PNAS(2006)} of {\em rhodobacter sphaeroides\/}, and in 
the reaction center \cite{Reaction center van Grondelle, Reaction center
Jennifer} of the Photosystem II, long-lived oscillations have been
experimentally observed at low (77 K) and room temperatures (300 K), indicating 
a strong vibronic coupling in these systems. 

To analyze the experimental findings in such large and complex
photoactive molecular complexes, a thorough 
comparison with theoretical calculations is essential, in order to arrive 
at a reliable interpretation of the measured 2D spectra. Since it is 
a difficult and computationally demanding task to determine 2D optical 
spectra, often only the population dynamics of exciton states is 
calculated. For the FMO complex, 
a rather small light-harvesting complex, the hierarchical equations of
motion \cite{Proc Natl Acad Sci USA 106 17255 (2009)} were applied and
quantum oscillations were observed on the time scale of the 2D experiments
employing an environmental Debye model spectral density with rather small
reorganization energy \cite{JCP 139 235102(2013)}. Employing the 
numerically exact quasiadiabatic propagator path integral (QUAPI) allowed
to use a more realistic measured environmental spectral density. However, this 
resulted in a decay of the decoherence faster than experimentally 
observed \cite{Phys Rev E 84
041926 (2011), Chem Phys Lett 478 234 (2009)}. This spectral density
could, more recently, be employed to calculate the 2D spectra of FMO with the 
hierarchy equation \cite{Tobias HEOM} and a reasonable agreement between
theory and experiment could be achieved.  
The calculations of QUAPI and the hierarchical equations of motion 
treated the coupling of the complex to environmental
fluctuations numerically exactly. However, the computational effort is 
immense, which makes the simulation of larger light-harvesting 
molecular complexes (which contain, typically, dozens to 
hundreds of excitonic subunits) virtually impossible. The need for a 
highly efficient numerical tool to calculate 2D optical spectra of 
large molecular complexes with a reasonable numerical effort and a 
satisfactory accuracy still exists and it is expected to 
continue to increase. 

Given their complex molecular structures, for the calculation of 2D spectra 
of large light-harvestors, approximate schemes are usually inavoidable. 
Standard Redfield equations \cite{Adv Magn Reson (1965)}, which invoke 
a lowest-order Born and a Markovian approximation, are good at 
weak system-bath coupling, but fail for strong coupling. 
The regime of intermediate system-bath coupling as present for the exciton
dynamics in photoactive complexes \cite{Thomas Renger 2006} is
typically also not properly treated within Redfield equations \cite{J Chem
Phys 130 234110 (2009), NalbachJCP2010, PRE 88 062719 (2015)}. Thus, the {\it modified\/} Redfield
theory (MRT) has been widely used for the description of energy transfer
processes of large molecules \cite{J Chem Phys 108 7763
(1998), Chem Phys 275 355 (2002), New J Phys 15 075013 (2013), J Phys Chem
B 108 7445 (2004), J Phys Chem Lett 4 2577 (2013), J Phys Chem B 109 10542
(2005), J Phys Chem A 117 34 (2013)}. In this approach, that contribution
of the system-bath coupling Hamiltonian, which is diagonal in the eigenbasis of the
system, is included fully, while a second-order perturbative
approximation is used for the off-diagonal coupling terms. The equation of motion 
of the MRT includes a population transfer within the reduced density matrix, but 
the accompanying population-transfer induced dephasing is neglected. 
The accuracy of the MRT in view of the dynamics of the reduced density matrix 
has been analyzed in detail \cite{JCP 142 034109(2015)}. Moreover, MRT  has been 
shown to have a somewhat wider range of applicability when compared to both 
the original Redfield and F\"orster theory \cite{Chem Phys 275 355 (2002)}.  
Also, linear absorption spectra for an ensemble of B850 rings have been
determined which shows that MRT includes non-Markovian effects which
clearly show up in the high-energy part of the static absorption
lineshapes \cite{J Chem Phys 124 084903 (2006)}. 
Different energy transfer components of LHCII trimer and phycoerythrin
545 have been revealed using MRT by simultaneous quantitative fits of the
absorption, linear dichroism, steady-states fluorescence spectra, and
transient absorption kinetics upon excitation at different wavelengths
\cite{Phys Chem Chem Phys 8 793 (2006)}. 

A more refined description of the quantum dissipative exciton dynamics 
is achieved in this work upon observing that the population-transfer induced 
electronic dephasing can be efficiently included in the quantum master equation. 
The off-diagonal terms in the quantum master equation now include 
the decoherence of excited states and electronic dephasing between 
ground and excited states by exploiting the relation 
$1/T_{2}=1/2T_{1}+1/T^{*}_{2}$ to estimate the different
contributions to the dephasing rate, where $T_{2}$ is the transverse relaxation
time and $T_{1}$, $T^{*}_{2}$ are the longitudinal relaxation time and
pure dephasing time, respectively. While working out the details with the results 
reported in this paper, this extended quantum master equation 
has also been independently put forward very recently 
in Ref.\ \cite{CP 447 46(2015)} and has been named the coherent modified Redfield 
theory (CMRT). To avoid confusion, we use to this nomenclature also here. 

For calculating 2D photon-echo spectra, essentially two different approaches 
are available. On the one hand, the response to the sequence of applied laser pulses can 
be calculated by evaluating the third-order optical response function 
\cite{ShaulBook}. Modified Redfield theory was successfully applied to 
simulate the 2D spectra of the double-ring LH2 aggregate of purple 
bacteria including both the B800 and the B850 ring \cite{JPCB 118
7533(2014)}. Comparing experimentally measured and theoretically calculated 
results of 2D spectra revealed that excitation energy transfer through
the LHCII happens on three time scales: sub-100fs relaxation through
spatially overlapping states, several hundred femtosecond transfer between
nearby chlorophylls, and picosecond energy transfer steps between layers
of pigments \cite{J Phys Chem B 113 15352 (2009)}. More recently, 2D
spectra of the reaction center in the Photosystem II were calculated with MRT
at low temperature, and the charge separation process was investigated by
MRT including charge transfer states in the model \cite{J Chem Phys 134
174504 (2011), New J Phys 15 075013 (2013)}. 

An alternative approach to calculate 2D optical spectra, which is especially useful when 
finite durations of the laser pulses as well as pulse 
overlap effects are taken into account, is the equation
of motion-phase matching approach (PMA) \cite{J Chem Phys 123 164112 (2005)}. 
Using the PMA in combination with the conventional Redfield equations, 
2D spectra of a single FMO subunit were studied, and the signature
of energy transfer was revealed by well-resolved peaks in the simulation
with adjustable pure dephasing of exciton states \cite{JCP 132
014501(2010)}. 

Although MRT is used to tackle many different problems in the 
study energy transport in photosynthetic complexes, 
no investigation of its reliability in calculating
nonlinear and, specifically, 2D optical spectra is at hand. In this work, we first verify 
the CMRT approach by comparing the population dynamics of FMO exciton states with 
numerically exact results of the QUAPI approach. In addition, we combine the 
CMRT with the PMA to calculate 2D photon-echo spectra for a simple dimer model. 
Again, the results of CMRT+PMA are benchmarked against numerically exact results of 
the QUAPI approach. For the long-time steady state
dynamics, the CMRT+PMA and QUAPI simulations show differences for intermediate and strong
system-bath coupling. However, for intermediate coupling, as it is typical in
photosynthetic complexes, the short time dynamics including dephasing times
and coherent beating frequencies are well described by CMRT+PMA. Hence, 
an efficient numerical scheme to calculate 2D photon-echo spectra with 
a reasonable computational effort is available. 

The remainder of this paper is organized in the following way. In the next
Section \ref{sec.model}, we briefly introduce the model of the FMO complex, 
for we compare the performance of the the CMRT+PMA 
in calculating a non-trivial population dynamics. Additionally, the
dimer model is introduced for which we compare results of 2D photon echo
spectra. A brief description of the CMRT+PMA and QUAPI for calculating the
reduced density matrix and 2D spectra is given in Section \ref{sec.tech}. 
In Section \ref{sec.res}, we give the results of the comparison, and a
thorough discussion is appended in Section \ref{sec.disc}, before we finish with a
conclusion.

%%%%%%%%%%%%%%%%%%%%%%%%%%%%%%%%%%%%%%%%%%%%%%%%%%%%%%%%%% Master equation
%%%%of MRT
%%%%%%%%%%%%%%%%%%%%%%%%%%%%%%%%%%%%%%%%%%%%%%%%%%%%%%%%%%

\section{Model} \label{sec.model}

In the framework of open quantum systems the Hamiltonian of the complete
system $H$ can be decomposed as four parts
\begin{eqnarray}\label{eq:full-Hamiltonian}
\begin{aligned}
 H &= H_{S}+H_{SB}+H_{B}+H_{ren},\\
 H_{S} &=
\sum^{N}_{m=1}\epsilon_{m}a^{\dagger}_{m}a_{m}+\sum^{N}_{m=1}\sum_{n<m}J_{
nm}(a^{\dagger}_{m}a_{n}+a^{\dagger}_{n}a_{m}),\\
 H_{B} &=
\sum^{N}_{m=1}\sum^{N^{m}_{b}}_{j=1}(\frac{p^{2}_{mj}}{2}+\frac{1}{2}
\omega_{mj}x^{2}_{mj}),
\end{aligned}
\end{eqnarray}
where $\epsilon_{m}$ is the on-site transition energy and $J_{nm}$ is the
intermolecular coupling. $N$ is the total number of monomers. $N^{m}_{b}$
is the number of bath modes coupled to molecule $m$, which we will take to
be infinity. $x_{mj}$ and $p_{mj}$ are the mass weighted position and
momentum of the $j$th harmonic oscillator bath mode with frequency
$\omega_{mj}$. The interaction term $H_{SB}=\sum_{m}K_{m}\Phi_m(x)$
induces the coupling between system and bath. It is assumed to be
separable such that $K_{m}$ only acts on the system subspace and
$\Phi_m(x)$ only on the bath degrees of freedom. In the following we
further assume a linear relation between bath coordinates and the system.
The system-bath interaction is then given as
\begin{eqnarray}\label{eq:interaction Hamiltonian}
\begin{aligned}
 H_{SB}&=\sum_{m}K_{m}\sum_{j}c_{mj}x_{mj}\, , %,\\
% &=\sum^{N}_{m=1}a^{\dagger}_{m}a_{m}\sum^{N_{b}}_{j=1}c_{mj}x_{mj},
\end{aligned}
\end{eqnarray}
and we furthermore restrict our considerations to pure electronic
dephasing only, i.e. $K_{m}=a^{\dagger}_{m}a_{m}$. The renormalization
term is 
\begin{eqnarray}\label{eq:renormalization}
\begin{aligned}
 H_{ren}=\sum_{m}\sum_{j}K^{2}_{m}\frac{c^{2}_{mj}}{m_{mj}\omega^{2}_{mj}}
\, .
\end{aligned}
\end{eqnarray}
This term compensates for artificial shifts of the system frequencies due
to the system-bath interaction. 

The influence of the bath is fully described by its bath spectral density
\be \label{specdens}
J_m(\omega) = \pi\sum_j \frac{c_{mj}}{2m_{mj}\omega_{mj}}\delta(\omega-
\omega_{mj}) = 2\lambda\frac{\omega \gamma}{\omega^2+\gamma^2}
\ee
with reorganization energy $\lambda$ and high-frequency cut off $\gamma$.
We assume the bath at each monomer to be independent (no cross-correlation
between baths \cite{NalbachNJP2010}) but with identical spectra, i.e.
$J_n(\omega)=J_m(\omega)$. 

Laser pulses acting on the exciton complex result in the addition of a
system-field interaction, i.e. $H_S\rightarrow H_S+F(t)$, which is defined
within the dipole approximation according to 
\begin{eqnarray}
\begin{aligned}
 F(t)=-X\hat{E}(t)+H.c.,
\end{aligned}
\end{eqnarray}
with the electric field of the laser pulse $\hat{E}(t)$ and the electronic
transition dipole operator $\boldsymbol{\mu}$ of the exciton system
\begin{eqnarray}
\begin{aligned}
 \boldsymbol{\mu}=X+X^{\dagger} \with X =
\sum^{N}_{m=1}\boldsymbol{\mu}_{m} a_m .
\end{aligned}
\end{eqnarray}
$\boldsymbol{\mu}_{m}$ determines the dipole strength and direction of the
$m$th monomer.

\subsection{Dimer}

The dimer is modeled by two monomers with site energies $\epsilon_{1}=-50$ 
cm$^{-1}=-\epsilon_{2}$ and a coupling $J=150$ cm$^{-1}$.
The two dipole moments are considered to be perpendicular to each other,
i.e. $\boldsymbol{\mu}_{1}\perp \boldsymbol{\mu}_{2}$. For the bath
spectral density as specified in Eq.\ (\ref{specdens}), we choose  
$\lambda= 50$ cm$^{-1}$ and $100$ cm$^{-1}$, and set $\gamma=
100$ cm$^{-1}$. 

\subsection{FMO}

We model a monomer subunit of the FMO trimer by including seven
bacteriochlorophyll {\it a} molecular sites. The single excitation
subspace is described by a Hamiltonian
\be\label{eq:Hamiltonian}
H_{F} ={\scriptsize \begin{pmatrix}
           240 & -87.7 &  5.5 & -5.9  &  6.7  & -13.7 & -9.9  \\
               &  315  & 30.8 &  8.2  &  0.7  &  11.8 &  4.3  \\
               &       &   0  & -53.5 & -2.2  & -9.6  &  6.0  \\
               &       &      &  130  & -70.7 & -17.0 & -63.3 \\
               &       &      &       &  285  &  81.1 & -1.3  \\
               &       &      &       &       &  435  & 39.7  \\
               &       &      &       &       &       &  245 
\end{pmatrix} }
\ee
in units of cm$^{-1}$, where we use the site energies and dipolar couplings
determined by Adolphs and Renger \cite{AdolphsRengerBJ2006} for the FMO
complex of {\it Chlorobium tepidum}. Bath parameters are chosen as
$\lambda$=35 cm$^{-1}$ and $\gamma$=53 cm$^{-1}$ following Ref.\cite{Proc
Natl Acad Sci USA 106 17255 (2009)}.

\section{\label{sec.tech}  Methods}

\subsection{\label{subsec: tdeRME}Coherent modified
Redfield theory (CMRT)}

The CMRT can be derived from the Nakajima-Zwanzig equation \cite{J Chem
Phys 124 084903 (2006)} using a scheme for the separation of the total
Hamiltonian which does not treat the whole system-bath interaction term
$H_{SB}$ perturbatively \cite{J Chem Phys 130 234110 (2009)}\cite{J Chem
Phys 111 3365 (1999)}. Instead, the Hamiltonian is separated according to
\begin{eqnarray}\label{eq:Hamiltonian of MRT}
\begin{aligned}
 H_{0}&=H_{S}+H_{B}+\sum_{\mu}\ket{\mu}\bra{\mu}H_{SB}\ket{\mu}\bra{\mu},
\\
 H'&=\sum_{\mu \nu, \mu \neq
\nu}\ket{\mu}\bra{\mu}H_{SB}\ket{\nu}\bra{\nu},
\end{aligned}
\end{eqnarray}
where $\ket{\mu}$ are eigenstates of $H_{S}$ and $H'$ is the off-diagonal
term of the system-bath interaction part in the exciton basis. In this basis,
$H_{0}$ is diagonal and the matrix elements read
\begin{eqnarray}\label{eq:system Hamiltonian}
\begin{aligned}
 \bra{\mu}H_{0}\ket{\mu}=\epsilon_{\mu}-\lambda_{\mu \mu \mu
\mu}+H_{B}(\mu),
\end{aligned}
\end{eqnarray}
where $\epsilon_{\mu}$ is the $\mu$th excitonic level of the 
system Hamiltonian and 
\begin{eqnarray}\label{eq:reorganization energy}
%\begin{aligned}
 \lambda_{\mu \nu
\mu'\nu'}&=&\sum_{m}\bra{\mu}K_{m}\ket{\nu}\bra{\mu'}K_{m}\ket{\nu'}\sum_{j}
\frac{c^{2}_{mj}}{2m_{mj}\omega^{2}_{mj}} \nonumber \\
%\end{aligned}
\end{eqnarray}
is the weighted reorganization energy. Moreover, 
\begin{eqnarray}\label{eq:bath}
%\begin{aligned}
 H_B(\mu) &=& \frac{1}{2}\sum_{\xi}\left[\frac{p^{2}_{\xi}}{m_{\xi}} +
m_{\xi}\omega^{2}_{\xi}\left(x_{\xi}+\sum_{k}\frac{\bra{\mu}K_k\ket{\mu}}{
m_{\xi}\omega^{2}_{\xi}}\right)^{2}\right] \nonumber \\
%\end{aligned}
\end{eqnarray}
describes a bath of harmonic oscillators with mass $m_{\xi}$, frequency
$\omega_{\xi}$ and momentum $p_{\xi}$, shifted due to the coupling
with the exciton state $\ket{\mu}$. 

In addition to the redefinition of the system and the bath
Hamiltonian, one has to define a different type of projection operator
which only projects on the diagonal part of the system density matrix in
the eigenstate basis. This is achieved by 
\begin{eqnarray}\label{eq:projectors}
\begin{aligned}
 \widetilde{P}&=\sum^{N}_{\mu=0}P_{\mu} \with
 P_{\mu}\cdot&=R^{\mu}_{eq} \mathrm{tr} \{\ket{\mu}\bra{\mu}\cdot\}\, ,
\end{aligned}
\end{eqnarray}
where $P_{\mu}$ is the projector onto the $\mu$th excitonic state and
$R^{\mu}_{eq}=\exp(-\beta H_{B}(\mu)) / Z^{\mu}_{eq}$ is the equilibrium density matrix
of the bath when the system is in the excitonic state $\ket{\mu}$.  Here, 
$Z^{\mu}_{eq}=\rm tr \exp(-\beta H_{B}(\mu))$ with $\beta=1/(k_B T)$ and $T$
being the temperature. 

Inserting these definitions into the Nakajima-Zwanzig equation,
determining $H'$ up to second order and invoking the time-dependent
population transfer rate, one obtains an equation of motion for the
population transfer terms in the form 
\begin{eqnarray}\label{eq:population}
 \frac{\partial}{\partial
t}\rho_{\mu\mu}(t)&=&\sum_{\nu\neq\mu}[R_{\mu\mu\nu\nu}(t)\rho_{\nu\nu}(t)-R_{
\nu\nu\mu\mu}(t)\rho_{\mu\mu}(t)], \nonumber \\
\end{eqnarray}
with the population transfer rates  \cite{J Chem Phys
108 7763 (1998)} 
\begin{eqnarray}\label{eq:MRT Tensor}
\begin{aligned}
 &R_{\mu\mu\nu\nu}(t)=2 \, {\rm Re }\int^{t}_{0}d\tau \mathrm{tr}
\{\ket{\nu}\bra{\nu}\exp(-iH_{0}\tau)H'\ket{\mu}\bra{\mu}\\
                 & \times R^{\mu}_{eq}\exp(iH_{0}\tau)H'\},\\
                 &=2\, {\rm Re }\int^{t}_{0}d\tau
\exp[-i\omega_{\mu\nu}\tau-g_{\mu\mu\mu\mu}(\tau)-g_{\nu\nu\nu\nu}
(\tau)\\&+g_{\nu\nu\mu\mu}(\tau)+g_{\mu\mu\nu\nu}(\tau)-2i(\lambda_{
\nu\nu\nu\nu}-\lambda_{\mu\mu\nu\nu})\tau]\\
&\times
\{\ddot{g}_{\mu\nu\nu\mu}(\tau)-[\dot{g}_{\nu\mu\nu\nu}(\tau)-\dot{g}_{
\nu\mu\mu\mu}(\tau)+2i\lambda_{\nu\mu\nu\nu}]\\
&[\dot{g}_{\nu\nu\mu\nu}(\tau)-\dot{g}_{\mu\mu\mu\nu}(\tau)+2i\lambda_{
\mu\nu\nu\nu}]\}.
\end{aligned}
\end{eqnarray}
Here, $\omega_{\mu\nu}=\epsilon_{\mu}-\epsilon_{\nu}$. The lineshape function
$g_{\mu\nu\mu'\nu'}(t)$ can be written as the
two-time integral of the bath correlation function according to 
\begin{eqnarray}\label{eq:lineshape function}
\begin{aligned}
 g_{\mu\nu\mu'\nu'}(t)&=\sum_{k}\bra{\mu}K_{k}\ket{\nu}\bra{\mu'}K_{k}\ket
{\nu'}\int^{t}_{0}d\tau\int^{\tau}_{0}d\tau'C(\tau'),\\
 \mbox{with}\;
C(t)&=\int^{\infty}_{-\infty}\frac{d\omega}{\pi}J(\omega)\frac{e^{i\omega
t}}{e^{\beta\omega}-1}.
\end{aligned}
\end{eqnarray}
To obtain Eq. (\ref{eq:MRT Tensor}), we have used the cumulant expansion
 up to second order in the system-bath coupling and have taken the independent
bath model into account. The absorption lineshape within the CMRT is given by 
\begin{eqnarray}\label{eq:absorption lineshape}
\begin{aligned}
 I(\omega)&={\rm Re }\sum_{\mu}d_{\mu}\int^{\infty}_{0}dt\exp[i(\omega-\omega_{
\mu 0})t-g_{\mu\mu\mu\mu}(t)\\
&-\frac{1}{2}\sum_{\nu\neq\mu}\int^{t}_{0}R_{\mu\mu\nu\nu}(\tau)].
\end{aligned}
\end{eqnarray}
as detailed in Ref.\ \cite{J Chem Phys 124 084903 (2006)}.

Up to this point, Eq.\ (\ref{eq:population}) constitutes the modified Redfield
theory, as developed and applied in Refs.\ \cite{J Chem Phys 108 7763
(1998), Chem Phys 275 355 (2002), New J Phys 15 075013 (2013), J Phys Chem
B 108 7445 (2004), J Phys Chem Lett 4 2577 (2013), J Phys Chem B 109 10542
(2005), J Phys Chem A 117 34 (2013)}. Based on the
population transfer term in Eq.\ (\ref{eq:population}), we extend the 
quantum master equation by including also the coherence (or, off-diagonal) terms
of the reduced density matrix. The resulting coherent modified Redfield quantum
master equation now reads 
\begin{eqnarray}\label{eq:diagonal population}
 \frac{\partial}{\partial t}\rho(t)&=&-i[H+F(t),\rho(t)]-\Re\{\rho(t)\}\, , 
\end{eqnarray}
where $F(t)$ is the time-dependent system-field interaction term. 

The relaxation and dephasing operator $\Re\{\rho(t)\}$ now also includes
diagonal and off-diagonal terms. The diagonal part of the relaxation
operator, which was desribed in Ref.\ \cite{J Luminescence 125 126 (2007)},
reads 
\begin{eqnarray}\label{eq:diagonal population}
\begin{aligned}
 \Re\{\rho(t)\}_{\mu\mu}&=\sum_{\nu\neq\mu}[R_{\mu\mu\nu\nu}(t)\rho_{
\nu\nu}-R_{\nu\nu\mu\mu}(t)\rho_{\mu\mu}]\, .
\end{aligned}
\end{eqnarray}
The off-diagonal terms $\Re\{\rho(t)\}_{\mu\nu}$ are now included in
order to describe decoherenece of excited states and
electronic dephasing between the ground and excited states. Here, we use 
an efficient way to obtain the associated rates by exploiting the
relation $1/T_{2}=1/2T_{1}+1/T^{*}_{2}$ to estimate the different
contributions to the dephasing rate. $T_{2}$ is the transverse relaxation
time, $T_{1}$, $T^{*}_{2}$ are the longitudinal relaxation time and
pure dephasing time, respectively \cite{book1}. In detail,
$1/T_{1}=\sum_{e\neq\mu}R_{\mu\mu ee}+\sum_{e\neq\nu}R_{\nu\nu ee}$ and
$1/T^{*}_{2}$ is given by the first derivative of lineshape function
${g}_{\mu\mu\nu\nu}(t)$. Therefore, the off-diagonal terms of the excited
states and between the ground and excited states can be written as 
\begin{eqnarray}\label{eq:off-diagonal population}
\begin{aligned}
 \Re\{\rho(t)\}_{\mu\nu}=&\left[\frac{1}{2}\left(\sum_{e\neq\mu}R_{\mu\mu
ee}(t)+\sum_{e\neq\nu}R_{\nu\nu ee}(t)\right)+\right.\\
                         &\left.\dot{g}_{\mu\mu\nu\nu}(t)
\right]\rho_{\mu\nu}(t),\\
 \Re\{\rho(t)\}_{\mu 0}=&\{\frac{1}{2}(\sum_{m\neq n}R_{m m n
n}(t)+\sum_{n\neq m}R_{n n m m}(t))+\\
                        &\dot{g}_{\mu\mu\mu\mu}(t)\}\rho_{\mu 0}(t).
\end{aligned}
\end{eqnarray}
This extended quantum master equation has also been independently put forward
very recently in Ref.\ \cite{CP 447 46(2015)} and has been named the coherent
modified Redfield theory (CMRT). It is an efficient, but approximate way to
take into account population transfer and dephasing on the same footing.

\subsection{QUAPI}

The quasiadiabatic propagator path integral \cite{MakriQUAPI1995a,
MakriQUAPI1995b} is a numerically exact
approach to determine the influence of environmental fluctuations on the
system dynamics within a open quantum systems approach. Specifically,
QUAPI determines the time dependent reduced statistical operator $\rho (t)$ of
the system. It is well established in the literature and we only briefly
summarize the central features in the following. The algorithm is
based on a symmetric Trotter splitting
of the short-time propagator ${\cal K}(t_{k + 1}, t_k)$ for the full
Hamiltonian into two parts, one depending on the system Hamiltonian, and
one involving the bath and the coupling term. The short-time propagator
determines the time evolution over a Trotter time slice $\delta t$. The
discrete time evolution becomes exact in the limit $\delta t \to 0$. For any
finite
$\delta t$, a finite Trotter error occurs which has to be eliminated by
choosing $\delta t$ small enough to achieve convergence. On the other
side, the environmental degrees of freedom generate correlations which are
non-local in time. For any finite temperature, these correlations decay on
a time scale denoted as the memory time scale. The QUAPI scheme defines an
augmented reduced density tensor, which lives on this full memory time window.
Then, an iteration scheme is established in order to extract the time evolution
of this object. All correlations are completely included over the finite
memory time  $\tau_{\rm mem} = K \delta t$ but are neglected for times
beyond $\tau_{\rm mem}$. One increases the memory parameter $K$ until
convergence is found. The two strategies to achieve convergence, i.e.,
minimize $\delta t$ but maximize $\tau_{\rm mem} = K \delta t$, are
naturally counter-current, but nevertheless convergent results can be
obtained in a wide range of parameters, including the cases presented in
this work.

\subsection{Two-dimensional electronic spectroscopy}

In the equation of motion-phasing matching approach (PMA), the polarization in
the photon-echo direction is calculated by simultaneously propagating
three auxiliary density matrices $\rho_{1}, \rho_{2}$ and $\rho_{3}$,
thereby employing also the rotating wave approximation \cite{J Chem Phys
123 164112 (2005)}. The time evolution equations are given by 
\begin{eqnarray}\label{eq:three main equation}
%\begin{aligned}
 \frac{\partial}{\partial
t}\rho_{1}(t)&=-i[H-V_{1}(t,t_{1})-V^{\dagger}_{2}(t,t_{2})-V^{\dagger}_{3
}(t,t_{3}),\rho_{1}(t)]\nonumber \\&-\Re\{\rho_{1}(t)\},\\
 \frac{\partial}{\partial
t}\rho_{2}(t)&=-i[H-V_{1}(t,t_{1})-V^{\dagger}_{2}(t,t_{2}),\rho_{2}(t)]
-\Re\{\rho_{2}(t)\},\nonumber\\
 \frac{\partial}{\partial
t}\rho_{3}(t)&=-i[H-V_{1}(t,t_{1})-V^{\dagger}_{3}(t,t_{3}),\rho_{3}(t)]
-\Re\{\rho_{3}(t)\}.\nonumber
%\end{aligned}
\end{eqnarray}
where
$V_{\alpha}(t,t_{\alpha})=XE_{\alpha}(t-t_{\alpha})=XE_{\alpha}(t-t_{
\alpha})e^{i\omega t}$ and
$E_{\alpha}(t-t_{\alpha})=\exp(-4\ln2(t-t_{\alpha})^{2}/\tau^{2}_{p})$,
$\tau_p$ is the pulse duration. To obtain the third-order 2D signal, the
polarization in the phase matching direction is evaluated as 
\begin{eqnarray}\label{eq:polarization signal}
%\begin{aligned}
 P_{PE}(t_{1},t_{2},t_{3})&=&e^{i\hat{k}_{s}\cdot \hat{r}}\langle
X(\rho_{1}(t)-\rho_{2}(t)-\rho_{3}(t))\rangle +c.c.\, , \nonumber \\
%\end{aligned}
\end{eqnarray}
where the bracket $\langle\ldots\rangle$ denotes the trace.
Experimentally, in the limit of ideal detection, the heterodyne photo echo
signal is proportional to the polarization $P_{PE}(t_{1}, t_{2}, t_{3},
t)$, where $t$ is the detection time. Therefore, the ideal total 2D
signal can be expressed as 
\begin{eqnarray}\label{eq:2D Fourier transform}
\begin{aligned}
 S_{T}(\omega_{\tau},T,\omega_{t}) \propto & \int^{\infty}_{-\infty}
d\tau\int^{\infty}_{-\infty} dt
e^{-i\omega_{\tau}\tau}e^{i\omega_{t}t}\\&\times iP_{PE}(\tau,T,t),
\end{aligned}
\end{eqnarray}
where $\tau$, $T$ and $t$ denote coherence time, population (waiting) time
and detection time, respectively, $\tau=t_{2}-t_{1}$, $T=t_{3}-t_{2}$. The
coherence time corresponds to a period in which the system is coherently
evolving after the first interaction with the optical field. The second
interaction with the field creates population states and the third
interaction recovers the coherence again. The Fourier transform in Eq.\ 
(\ref{eq:2D Fourier transform}) is always performed over the coherence
time $\tau$ and the detection time $t$. The corresponding frequencies
$\omega_{\tau}$, $\omega_{t}$ are often referred to as absorption and
emission frequencies, respectively. In addition, Gaussian laser pulses
have been assumed for a realistic detection scheme, which have the form
\begin{eqnarray}
\begin{aligned}
 \hat{E}(t)=\sum^{3}_{\alpha=1}Ae^{-4\ln 2(t-t_{\alpha})^{2}/\tau^{2}_{p}}e^
{i\omega t}e^{-i\hat{k}_{\alpha}\hat{r}}+c.c.
\end{aligned}
\end{eqnarray}
where $A$, $t_{\alpha}$, $\hat{k}_{\alpha}$, and $\omega$ are the
amplitude, envelop central time, wave vector and frequency of the pulses
and $\tau_{p}$ characterizes the pulse duration. Note that all the pulses
are assumed to have the same lineshape, carrier frequencies and durations
in this paper.

\section{Results}
\label{sec.res}

\subsection{Population dynamics of the FMO complex} 

\begin{figure}
\begin{center}
\includegraphics[width=7.5cm]{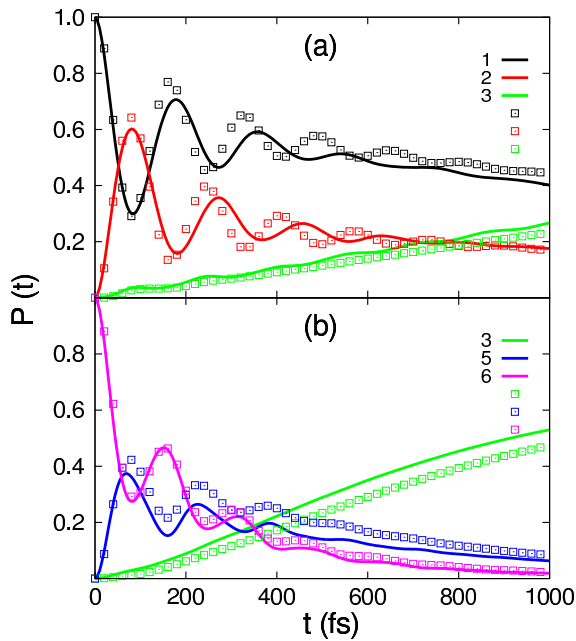}
\caption{\label{fig:FMO} Population dynamics of selected FMO sites. In a), the
population of sites $1$ (black), $2$ (red) and $3$ (green) with the initial
condition $\rho(1,1)=1$ is shown, while in (b) the populations of sites 
$3$ (green), $5$ (blue) and $6$ (magenta) with the initial
condition $\rho(6,6)=1$ is depicted (symbols: QUAPI, full lines:
CMRT) for the parameters as given in the text.}
\end{center}
\end{figure}

In order to verify the reliability of the CMRT, we present the
population dynamics of the FMO complex calculated by CMRT and compare the
results to
those obtained by the numerically exact QUAPI method. In Fig.\ 
\ref{fig:FMO}, the population dynamics of selected FMO sites is shown for
$T=77$ K for two different initial conditions. In Fig.\ \ref{fig:FMO}a), 
we assume the energy transfer to start from site 1. We monitor then the full
transfer which involves all seven FMO sites. For simplicity, we only show the
population dynamics of the sites 1, 2, and 3. Alternatively, the energy transfer
may be assumed to start from site 6, see Fig.\ \ref{fig:FMO}b). There,
we depict the population dynamics of the relevant sites 3, 5, and 6. We
observe that the oscillatory
behavior of the populations is captured by both approaches. Both
also yield the same decay rates and periods of oscillations. However,
a phase shift of the oscillations occurs between the CMRT and QUAPI
results. Energy transfer is believed to be related to the population of the
FMO site 3 (green symbols and lines) which has the lowest energy in the FMO
monomer. In our comparison, CMRT slightly overestimates the population
transfer efficiency towards site 3. All in all, the CMRT results for the FMO
exciton population dynamics are in good agreement to numerically exact
QUAPI results. Since the system-bath coupling parameters of
the FMO complex are typical for natural photosynthetic units, we conclude
that CMRT is a useful tool to study their exciton dynamics.

%%%%%%%%%%%%%%%%%%%%%%%%%%%%%%%%%%%%%%%%%%%%%%%%%%%
%%%%%%%%%%%%%%%%%%%%%%%%%%%%%%%%%%%%%%%%%%%%%%%%%%%

\subsection{Two-dimensional spectra of a dimer system} 

To obtain 2D spectra, we combine the CMRT next with the PMA. This constitutes a
very efficient approximate numerical tool whose reliability is assessed by a
comparison with 2D spectra obtained by QUAPI. Since 2D spectra involve extended
numerical calculations, QUAPI results are available only for small model systems
with present day hardware technology. For such a comparison, we present the
calculated results for the dimer model. It allows us to study energy transfer
and dephasing (homogeneous broadening) as building blocks of the exciton
dynamics in larger molecular compounds. It can still be treated by QUAPI
with reasonable numerical effort. 

\begin{figure}
\begin{center}
\includegraphics[width=8.5cm]{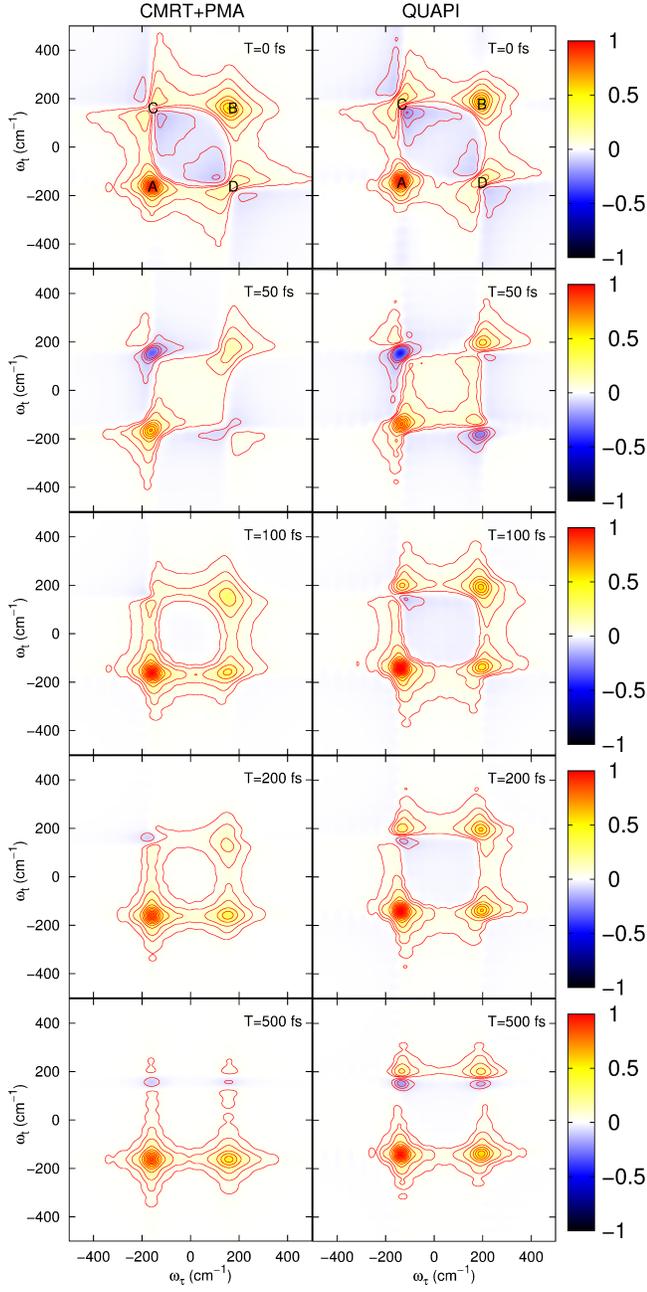}
\caption{\label{fig:2D dimer} Two-dimensional photon-echo spectra of a
dimer model calculated by CMRT+PMA (left) and QUAPI (right) for different
waiting times as indicated. The Debye spectral density was used for the
calculation with the parameters $\lambda=50$ cm$^{-1}$, $\gamma=100$ cm$^{-1}$
and the temperature was set to $T=77$ K.}
\end{center}
\end{figure}

Fig.\ \ref{fig:2D dimer} (left) shows 2D spectra of the dimer calculated by
CMRT+PMA for $\lambda$=50 cm$^{-1}$ and the other parameters as indicated
above. They are compared to QUAPI results (right column in Fig.\ \ref{fig:2D
dimer}) for waiting times $T=0$ fs, $50$ fs, $100$ fs and $500$ fs. These 2D
spectra show two diagonal peaks (labeled A, B) which correspond to the two
exciton states. Moreover, two cross peaks (labeled C and D) arise due to
the excitonic coupling between them. For the sake of simplicity and clarity of
the comparison, inhomogeneous broadening and the rotational averaging for
different laser polarizations and molecular orientations is not performed here.
Although this would be important to describe realistic experimental situations, 
the averaged results generally show smaller discrepancies (not shown).

At $T=0$ fs, the two results show the same profile for diagonal and cross
peaks and, indeed, the agreement is excellent. This shows that the CMRT 
 correctly models the coherence times and the system-bath correlations created
during the simulation. With increasing waiting time,
the same coherent dynamics is found for both the diagonal and the cross peaks
and even can be inspected by eye. However, some disagreement is observed at
long waiting time $T=500$ fs. The diagonal peak B in left figure (CMRT+PMA)
shows a somewhat reduced amplitude as compared to the right figure (QUAPI). 

\begin{figure}
\begin{center}
\includegraphics[width=6.5cm]{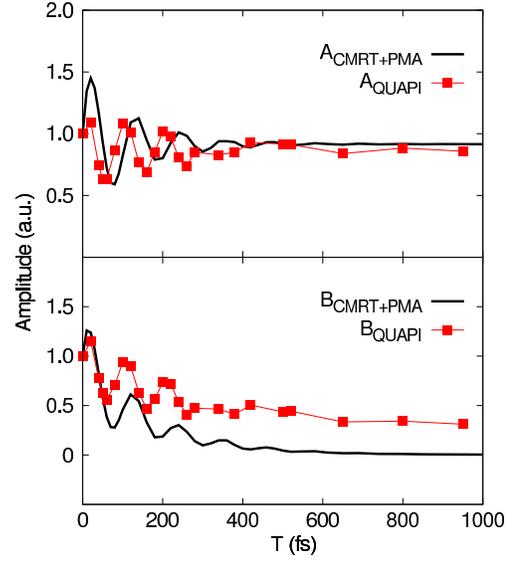}
\caption{\label{fig:diagonal_dynamics} Population dynamics of the labeled
diagonal peaks (A, B) extracted from the underlying sequence of 2D maps. The two
approaches yield the same oscillation period. The diagonal peak B
obatained from CMRT+PMA decays faster as compared to the QUAPI result. The
oscillation periods can be extracted by data fitting and are:  CMRT+PMA: 110fs,
QUAPI: 99fs.}
\end{center}
\end{figure}

\begin{figure}
\begin{center}
\includegraphics[width=6.5cm]{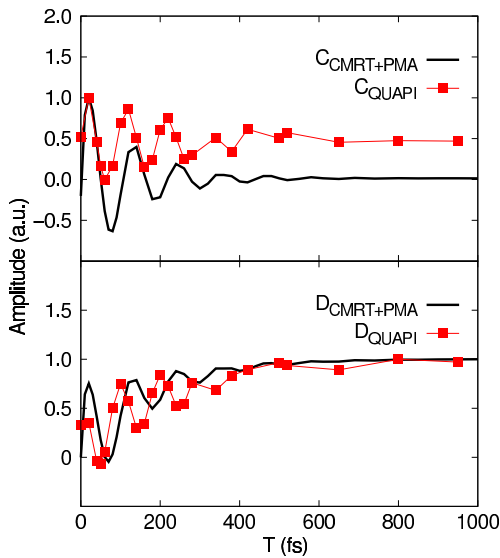}
\caption{\label{fig:anti-diagonal_dynamics} Coherent oscillations of
labeled cross peaks (C, D) extracted from 2D maps. Cross peak C obained by
CMRT+PMA shows the same oscillatory behavior, but with a somewhat smaller
amplitude.}
\end{center}
\end{figure}

For a more refined comparison, the amplitudes of the diagonal and cross peaks
(A, B, C and D) are plotted against the waiting time in Fig.\ 
\ref{fig:diagonal_dynamics} and Fig.\ \ref{fig:anti-diagonal_dynamics}. In 
Fig.\ \ref{fig:diagonal_dynamics}, the 
population dynamics of the diagonal peaks A (top) and B (bottom) calculated by
CMRT+PMA from $0$ to $1000$ fs is shown and compared to the QUAPI result.
We find that the CMRT+PMA provides reasonably accurate results for the
population transfer and the oscillation period. However, the amplitude of peak
B decays slightly faster in the approximate results as compared to the 
QUAPI data. Moreover, both yield different stationary states. In
addition, the phase of the oscillations is slightly shifted. For the
comparison of the cross peaks, the oscillatory behavior of peaks C and D is
plotted in Fig.\ \ref{fig:anti-diagonal_dynamics}. Cross peak C shows
a similar oscillatory behavior but the two approaches yield different stationary
states. Peak D shows a only slightly shifted phase of the
oscillatory behavior. Such a phase shift was also observed in the population
dynamics of the FMO complex shown above. The phase shift might be due to the
neglect of imaginary parts in the Redfield relaxation tensor. 

\begin{figure}
\begin{center}
\includegraphics[width=6.5cm]{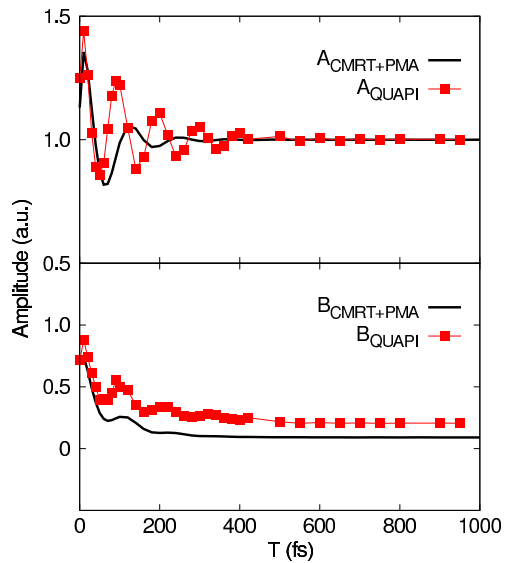}
\caption{\label{fig:diagonal_dynamics2} Amplitude of the diagonal
peaks A and B for a stronger system-bath coupling $\lambda$=100
cm$^{-1}$ (with $\gamma$=100 cm$^{-1}$ unchanged). CMRT+PMA calculations yield
 a faster decay (A) as compared to the QUAPI
result (decay rate extracted from a fit: CMRT+PMA: 81 fs, QUAPI: 146
fs).}
\end{center}
\end{figure}

\begin{figure}
\begin{center}
\includegraphics[width=6.5cm]{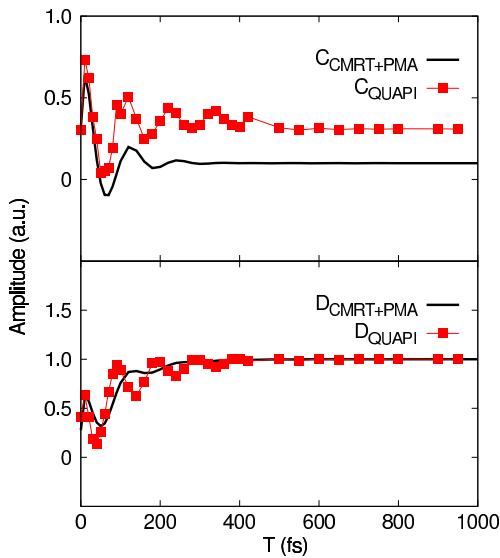}
\caption{\label{fig:anti-diagonal_dynamics2} Oscillations of the 
 cross peaks (C, D) with for a stronger system-bath interaction (same
parameters as in Fig.\ \ref{fig:diagonal_dynamics2}). The cross-peak
C calculated by CMRT+PMA yields a faster decay and a smaller amplitude 
(decay rates as obtained from a fit: CMRT+PMA: 62 fs, QUAPI: 177 fs).}
\end{center}
\end{figure}

In order to further assess the reliability of the CMRT+PMA, we
have repeated the calculations for a larger reorganization energy, i.e., 
for $\lambda$=100 cm$^{-1}$ (with $\gamma$=100 cm$^{-1}$ kept unchanged). 
2D spectra were again calculated by both approaches and the amplitude of the 
labeled peaks were extracted. Their time-dependence is plotted in Figs.\ 
\ref{fig:diagonal_dynamics2} and \ref{fig:anti-diagonal_dynamics2}. CMRT+PMA
still yields quantitative agreement with the QUAPI result except for the
behavior of the damping. The stronger
system-bath coupling results in faster damping (diagonal peak A) and
also in an increased difference between QUAPI and CMRT+PMA as compared to the
weaker coupling with $\lambda$=50 cm$^{-1}$.

%%%%%%%%%%%%%%%%%%%%%%%%%%%%%%%%%%%%%%%%%%%%%%%%%%%
%%%%%%%%%%%%%%%%%%%%%%%%%%%%%%%%%%%%%%%%%%%%%%%%%%%

%%%%%%%%%%%%%%%%%%%%%%%%%%%%%%%%%%%%%%%%%%%%%%%%%%%
%%%%%%%%%%%%%%%%%%%%%%%%%%%%%%%%%%%%%%%%%%%%%%%%%%%

%%%%%%%%%%%%%%%%%%%%%%%%%%%%%%%%%%%%%%%%%%%%%%%%%%%
%%%%%%%%%%%%%%%%%%%%%%%%%%%%%%%%%%%%%%%%%%%%%%%%%%%

\section{Discussion -- CMRT+PMA versus QUAPI} 
\label{sec.disc}

From the above comparison of the results obtained by both approaches, we
observe that the discrepancies found in the 2D calculations are more
pronounced than in the dynamics of the populations. Put differently, nonlinear
2D spectra  are more sensitive to assess the performance and reliability of
approximate theoretical approaches. In order to understand this, we point out 
two fundamental differences between 2D spectra and the population dynamics.
First, entanglement between the system and the bath leads to initial
correlations at the beginning of the waiting time window, which are absent in
the calculation of the population dynamics. Second, two-exciton states
contribute to the 2D
spectra during the detection time, and interference between positive and
negative peaks changes the observed amplitudes. This shows that one can
not understand the reliability of a method to simulate correct 2D
spectra by calculating population dynamics alone. Our current framework,
in which we use the combined CMRT+PMA and compare the results with QUAPI,
is well suited to show the performance of these methods in understanding 2D
spectra directly. 

In more detail, we have observed three noticeable discrepancies of the CMRT+PMA
as compared to QUAPI: i) shifted oscillation phase of peak intensities, ii) a 
slightly faster decay, and, iii) a different amplitude of peaks B and C for 
long waiting times. 

For the explanation of the shifted oscillation observed in 2D simulations 
of the CMRT+PMA, we need to notice that Eqs.\ \ref{eq:diagonal population},
\ref{eq:off-diagonal population} provide the analytic result for a monomer
(two-level system), and that this has been proven by comparing to QUAPI
\cite{MRT_QUAPI}. However, CMRT yields a shifted period for the dimer
model. The mismatch is mainly caused by the population transfer term
$R(t)$ since there is no population transfer term in the monomer model. In
this paper, the population transfer rate was calculated by the cumulant
expansion in Eq.\ \ref{eq:MRT Tensor} \cite{J Chem Phys 108 7763 (1998)}
and we only took the real part. It is well known that the imaginary part
dominates the phase of the oscillations \cite{JCP 130 234111(2009)}. So, most
likely, the shifted oscillation is mainly caused by the real-value
approximation of the population transfer rates.

Then, the population transfer term is also derived based on the
second-order perturbation approximation, which is one of the reasons for
the explanation of the slightly too fast decay of the oscillations found in 
CMRT calculations. Furthermore, the secular approximation was used to separate
the population dynamics and the dephasing process in Eqs.\ \ref{eq:diagonal
population}, \ref{eq:off-diagonal population}. This also contributes to
the discrepancy in the decay rate, since it neglects the interference
between population transfer and coherence dephasing. 

A relatively small amplitude of peak B and C was found in Fig.\ 
\ref{fig:diagonal_dynamics} and Fig.\ \ref{fig:anti-diagonal_dynamics} and
it also can be observed by eye in the 2D map for the long waiting time $T=500$ 
fs. We observe that peaks B and C are mainly formed by one positive (red)
peak and overlap with a negative (blue) peak in the 2D spectrum ($T=500$ 
fs). Therefore, the amplitude of those peaks mainly depends on the
overlap of two peaks. In the QUAPI result, the two peaks are clearly
separated with a larger spectral distance than in the CMRT result and this
leads to the larger amplitude of peaks B and C in the 2D spectrum calculated
with
QUAPI. It indicates that, besides the shifted oscillation and the faster decay
of the oscillation, CMRT does not properly account for the reorganization
energy by the heat bath (diagonal peaks show slightly different positions
in the 2D map: $-190$ cm$^{-1}$ and $190$ cm$^{-1}$ for CMRT and $-180$
cm$^{-1}$ and $200$ cm$^{-1}$ for QUAPI). In the CMRT, the reorganization
energy is included in the diagonal part of the Hamiltonian by Eq.\ 
\ref{eq:renormalization}, where it just brings in a shift of the
excitonic transition frequency $E_{\mu}$ by the renormalization term Eq.\ 
\ref{eq:renormalization} and does not affect the dynamics of the off-diagonal
terms in the Hamiltonian in Eq.\ (\ref{eq:Hamiltonian of MRT}).

%%%%%%%%%%%%%%%%%%%%%%%%%%%%%%%%%%%%%%%%%%%%%%%%%%%
%%%%%%%%%%%%%%%%%%%%%%%%%%%%%%%%%%%%%%%%%%%%%%%%%%%

%%%%%%%%%%%%%%%%%%%%%%%%%%%%%%%%%%%%%%%%%%%%%%%%%%%
%%%%%%%%%%%%%%%%%%%%%%%%%%%%%%%%%%%%%%%%%%%%%%%%%%% 

On the basis of a clear physical meaning (population transfer and dephasing
terms) and for the purpose of an efficient and fast calculation, the secular
approximation and the second-order perturbation theory were applied to construct
the CMRT. On the one hand, the secular approximation leads to a separation of
the
population dynamics and dephasing process and avoids any complicated
interaction terms between diagonal and off-diagonal parts in the equation
of motion. On the other hand, the second-order perturbation theory
simplifies the population transfers. It is possible to improve the
equation by including higher orders. However, this renders 
the equation considerably more complicated and requires more
computational resources for the
simulation and it is a priori unclear how much this improves the accuracy. 

%%%%%%%%%%%%%%%%%%%%%%%%%%%%%%%%%%%%%%%%%%%%%%%%%%%%%%%%%% Conclusions
%%%%%%%%%%%%%%%%%%%%%%%%%%%%%%%%%%%%%%%%%%%%%%%%%%%%%%%%%%
\section{\label{sec:conclusions} Conclusions}

In this paper we present the CMRT and compare it in more detail to the QUAPI
method by calculating the population dynamics of selected FMO exciton sites and
the 2D-spectrum of a model dimer.
We found that CMRT provides numerical reliable results as compared to
numerically exact QUAPI calculations for both the population dynamics and 2D
spectra, as long as the  reorganization energy is not too large compared to the
typical energy gap of the system. Most importantly, it requires smaller
computational efforts and orders of magnitude shorter calculation times. It 
provides us with an efficient approach to study the energy transfer in
super-large molecular complexes and to perform complicated 2D simulations. 

We found that the 2D profile calculated from CMRT+PMA agrees well with the
corresponding QUAPI results. For a quantitative comparison, the amplitudes of
diagonal and cross peaks were extracted from 2D maps and compared to those
calculated by QUAPI. Quantitative agreement was found. We observe some
discrepancies. In particular, oscillations are shifted, they decay slightly
faster, and positions of peaks are slightly shifted. This becomes more serious
if the reorganization energy is increased, which is also the case in Ref.\ 
\cite{MaMoix2, MaMoix1}.

The simulation protocol developed here can be used for arbitrary forms of
the spectral density. One can envision an approach where the CMRT method
is used to simulate super-large complexes, while numerically exact methods
such as QUAPI play a role in benchmarking the accuracy of the simulations
of smaller systems.

%%%%%%%%%%%%%%%%%%%%%%%%%%%%%%%%%%%%%%%%%%%%%%%%%%%%%%%%%%
%%%Acknowledgements
%%%%%%%%%%%%%%%%%%%%%%%%%%%%%%%%%%%%%%%%%%%%%%%%%%%%%%%%%%
\begin{acknowledgments} 
We acknowledge financial support by the Joachim-Herz-Stiftung, Hamburg
within the PIER Fellowship program (HGD) and by the excellence cluster "The
Hamburg Center for Ultrafast Imaging - Structure, Dynamics and Control of
Matter at the Atomic Scale" of the Deutsche Forschungsgemeinschaft. AGD 
was supported by a Marie Curie International Incoming Fellowship within
the 7th European Community Framework Programme. PN acknowledges financial
support by the DFG project NA394/2-1.
\end{acknowledgments}

%%%%%%%%%%%%%%%%%%%%%%%%%%%%%%%%%%%%%%%%%%%%%%%%%%%%%%%%%% Bibliography
%%%%%%%%%%%%%%%%%%%%%%%%%%%%%%%%%%%%%%%%%%%%%%%%%%%%%%%%%%
%\bibliographystyle{model1a-num-names}
%\bibliography{RefsFMO}

\begin{thebibliography}{69}
\bibitem {MRT_QUAPI} We compare the absorption lineshape calculated by CMRT
and QUAPI. They show exactly same result.
\end{thebibliography}

%%%%%%%%%%%%%%%%%%%%%%%%%%%%%%%%%%%%%%%%%%%%%%%%%%%%%%%%%%

\end{document}